\documentclass[a4paper,11pt]{article}
\usepackage{pos}
\usepackage{enumitem}
\usepackage{wrapfig}
\usepackage[utf8]{inputenc}
\usepackage{array,multirow,graphicx}

\title{A composition-informed search for large-scale anisotropy with the Pierre Auger Observatory}
\ShortTitle{A composition-informed search for large-scale anisotropy with the Auger Observatory}

\author*[a]{Geraldina Golup}

\affiliation[a]{CNEA/CONICET, Centro At\'omico Bariloche, Argentina.}

\affiliation[b]{Observatorio Pierre Auger, Av.\ San Mart{\'\i}n Norte 304, 5613 Malarg\"ue, Argentina\\
Full author list: \normalfont{\url{https://www.auger.org/archive/authors\_icrc\_2025.html}}}

\onbehalf{for the Pierre Auger Collaboration$^{b}$} 

\emailAdd{spokespersons@auger.org}

\abstract{The large-scale dipolar structure in the arrival directions of ultra-high-energy cosmic rays with energies above $8\,$EeV observed by the Pierre Auger Collaboration is a well-established anisotropy measurement. This anisotropy is understood to be of extragalactic origin, as the maximum of the dipolar component is located ${\sim}115^\circ$ away from the Galactic Center. Cosmic rays interact with background radiation and magnetized regions on their path from their sources to Earth. These interactions, which depend on the cosmic-ray energy, charge and mass composition, give rise to different horizons and deflections that are expected to lead to different anisotropies in the arrival directions of cosmic rays at Earth. The Auger Collaboration has determined that the mass composition of cosmic rays at ultra-high energies is mixed, becoming increasingly heavier as a function of energy. Thus, different dipole amplitudes are expected to be measured at a given energy when separating the data into composition-distinct subsets of lighter and heavier elements.

In this contribution, we investigate the composition signature on the large-scale anisotropy taking advantage of composition estimators obtained from the data gathered with the surface detector. A way of probing for composition signatures in anisotropy patterns is then to divide the data into subsets of ``lighter'' and ``heavier'' elements per energy bin. In a simulation library, we evaluate the possibility of measuring a separation in total dipole amplitude between two such populations of the measured dataset under a source-agnostic model. We present the results using two different composition estimators, one based on air-shower universality and one inferred with deep learning.}

\ConferenceLogo{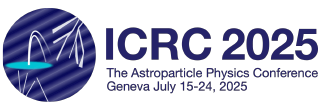}

\FullConference{%
39th International Cosmic Ray Conference (ICRC2025)\\
  15 - 24 July, 2025\\
  Geneva, Switzerland}

\begin{document}
\maketitle

\section{Introduction}
The observation of the modulation in right ascension in the arrival directions of ultra-high-energy cosmic rays (UHECRs) with energies above $8\,$EeV \cite{AugerLSA}, enabled by the high-quality data from the Pierre Auger Observatory \cite{AugerNIM2015}, is a milestone in cosmic ray physics. Its significance with the complete Phase 1 data set is $6.8\sigma$ \cite{AugerLSA24} and its direction, ${\sim}115^{\circ}$ from the Galactic center, is indicative of an extragalactic origin for cosmic rays above this energy threshold. 

Additionally, an approximately linear increase of the dipole amplitude with energy has been detected \cite{AugerLSA18}. This trend is expected to be a consequence of cosmic rays interactions with background radiation and magnetized regions on their path from their sources to Earth. First, cosmic rays suffer energy losses when interacting with background radiation, leading to a decrease of the horizon of cosmic-ray sources and an increase of the dipole amplitude due to the growing relative contribution of nearby sources, whose distribution is more inhomogeneous. Second, the increase of the UHECR magnetic mean rigidity, from ${\approx}4\,$EV to ${\approx}8\,$EV between $10\,$EeV and $100\,$EeV, also results in a growth of the dipole amplitude with energy. These effects result in varying horizons and deflections depending on the composition and energy of the cosmic rays, which in turn produce different dipolar amplitudes\footnote{In the alternative scenario that there is a dominant source at the highest energies, diffusion in the extragalactic magnetic field  leads to a dipole that increases with rigidity, and thus also a total dipole increasing with energy for the measured composition.}. In this work, we study the possibility of detecting a difference on the dipolar amplitudes for distinct composition-selected subsets.

The results from the Pierre Auger Observatory from the measurements of the depth of the shower maximum with the fluorescence detector indicate that the composition of UHECRs is mixed and that it becomes heavier with increasing energy \cite{augermass_frac, combined_fit}. In this work, we take advantage of composition estimators derived on an event-by-event basis using the surface detector of the Pierre Auger Observatory. In a Universality-based approach, the depth of the shower maximum, $X_{\rm max}^{\rm Univ}$, and the number of muons, $R_\mu^{\rm Univ}$, are reconstructed using a model grounded in air-shower universality  \cite{Universality}. These observables are then used to infer a proxy of the logarithm of the atomic mass, $\ln A^{\rm Univ}$, through a parameterization obtained from simulations \cite{max_lnA}, based on the EPOS-LHC hadronic interaction model \cite{EPOS}, with the estimate being primarily driven by $X_{\rm max}^{\rm Univ}$. Another mass estimator we consider is the depth of the shower maximum, $X_{\rm max}^{\rm DNN}$, reconstructed using deep neural networks (DNN) \cite{Aixnet}. While these estimators are not direct measurements, as the depth of the shower maximum is with the fluorescence detector, they have a factor ${\approx}10$ much larger statistics with respect to the data set from the fluorescence detector. The resolutions in $X_{\rm max}$ are ${\approx}40\,$g$\,$cm$^{-2}$ and 25$\,$g$\,$cm$^{-2}$, for Universality-based and DNN-based reconstructions, respectively. For comparison, the resolution of $X_{\rm max}^{\rm FD}$ measured with the fluorescence detector is ${\approx}15\,$g$\,$cm$^{-2}$ \cite{fd}. 

The size of the data set assumed in this work corresponds to that of Phase 1 of the Pierre Auger Observatory, as in \cite{AugerLSA24}, but only includes events with zenith angle smaller than $60^\circ$ (rather than $80^\circ$ as in \cite{AugerLSA24}), given that the mass estimators considered here can only be reconstructed below this zenith threshold. The corresponding total exposure is $92{,}000\,$km$^2\,$sr$\,$yr.

\section{Model for a rigidity-dependent dipole}

\begin{figure}[!ht]
   \centering  \includegraphics[width=.6\textwidth] {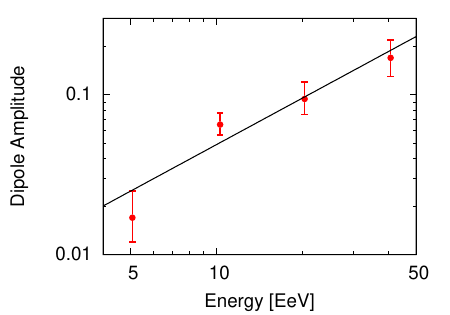}
\caption{
Evolution of the dipole amplitude with energy, for the energy bins: (4-8, 8-16, 16-32, $\ge32$)\,EeV \cite{AugerLSA24}.}
    \label{fig:LSA_3Ddipole}
\end{figure}

For energies above $4\,$EeV, the amplitude of the dipole $d$ is observed to increase with energy. As show in Fig.~\ref{fig:LSA_3Ddipole}, this behavior can be described with a power law, $d(E)=d_{10}\times \left( \frac{E}{10\ {\rm EeV}} \right)^\beta$, with the following parameters: $d_{10}=0.049 \pm 0.009$ and $\beta=0.97 \pm 0.21$ \cite{AugerLSA24}. Building on this result, for this work, we model the dependence of the dipole amplitude $d$ with charge $eZ$ and energy $E$, i.e. rigidity $E/Ze$, as \cite{Emily}
\begin{equation}
d(E,Z)= d_R \left (\frac{E/{\rm EeV}}{Ze} \right)^{\beta_R}
\end{equation}
where $d_R$ and $\beta_R$ are parameters that are derived using a simulation library, reproducing the measured dipole amplitude, $d(E)$. This behavior could arise either from models with a locally dominant source whose events diffuse through extragalactic magnetic fields, or from models where numerous inhomogeneously distributed sources contribute to the flux. We consider an upper limit of $d_{\rm max}=1$, corresponding to the scenario in which many sources contribute to the dipolar anisotropy. Moreover, to avoid making assumptions on the specific sources, we assume that the dipole direction is the same for all nuclei.  

\begin{figure}[!ht]
    \centering \includegraphics[width=.99\textwidth]{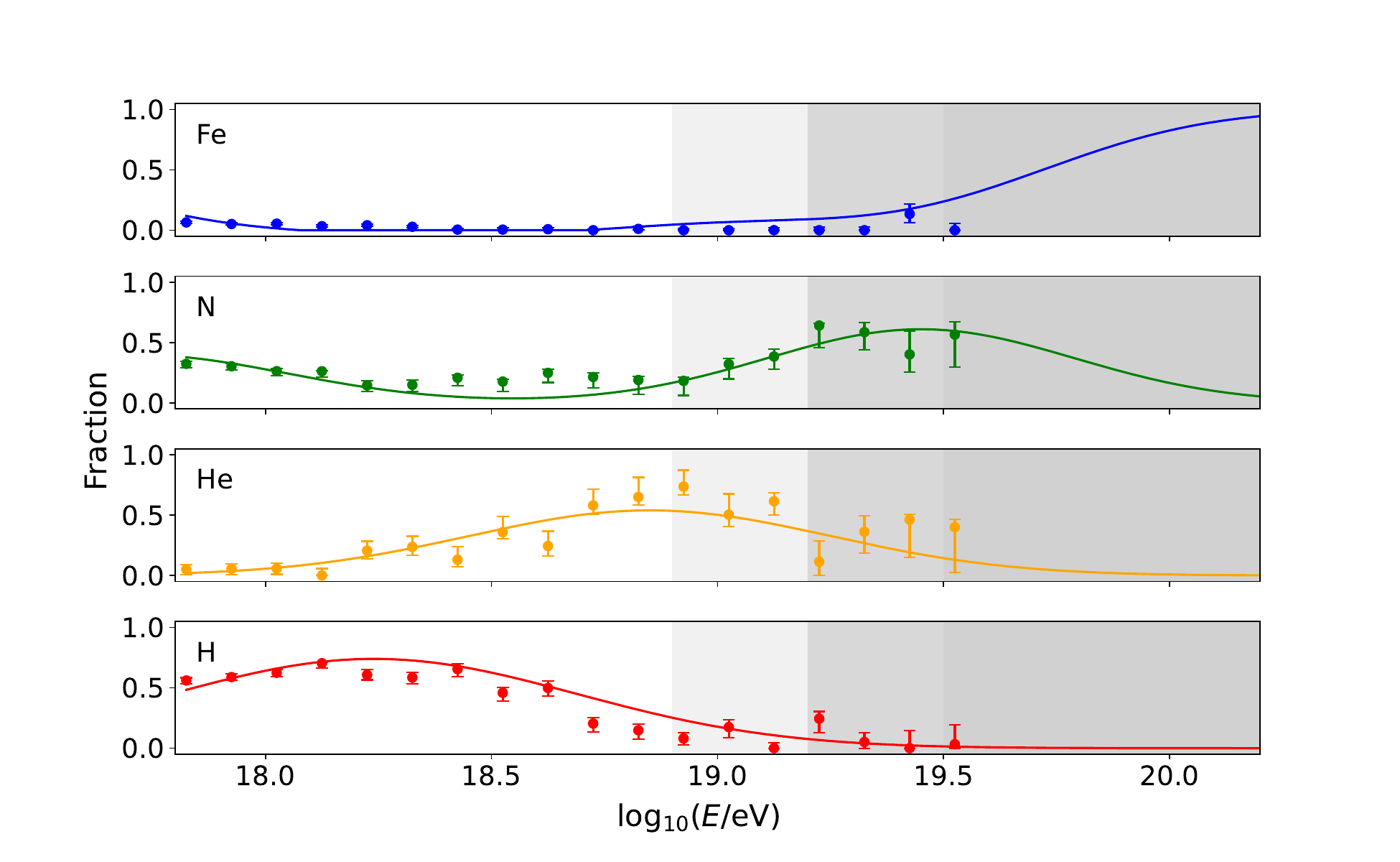}
\caption{Results of the fractional mass composition of the UHECR derived from $X_{\rm max}^{\rm FD}$ distributions using EPOS-LHC predictions for proton (red), helium (orange), nitrogen (green), and iron (blue), from  \cite{fractions}. The continuous lines are the model predictions assuming Gaussian functions with the energy of the position of the maximum ordered proportionally to $A$. The gray bands are included to indicate the energy bins considered in this work: (8-16, 16-32, $\ge32$)\,EeV.}
    \label{fig:AugerMix}
\end{figure}

The simulation library we use accounts for the detector response of both mass estimators, $\ln A^{\rm Univ}$ and $X_{\rm max}^{\rm DNN}$. The simulated energy spectrum follows the one detected with the Auger Observatory. The composition model used in this work, shown in Fig. 2, reproduces the mass composition fractions obtained in \cite{fractions}, where a fit of the $X_{\rm max}^{\rm FD}$ distributions is done, considering the EPOS-LHC hadronic interaction model, for a combination of four elemental groups: proton, helium, nitrogen and iron. In this work, the energy-dependent fraction of each nuclear species is described by a Gaussian curve, fitted to the composition fractions obtained from data. The peak energy of each Gaussian is set proportionally to the atomic mass number $A$ of each species, relative to the energy peak determined for protons\footnote{There is a second maximum for nitrogen at an energy of $\log (E/{\rm eV}) \approx 17.7$, which is not relevant in this analysis.}. This approach allows the extrapolation of the composition predictions to higher energies than those available with the statistics of the data set from the fluorescence detector.

Using this model, the best fit to data is $d_R=0.0018$ and $\beta_R=2.1$, as shown in the top panel of Fig.~3 \cite{Emily}. We also display in the bottom panel of Fig. 3, the total dipole amplitude along with the contributions from each component, computed according to Eq. (1) using the best fit parameters. We did not fit the $(4-8)\,$EeV bin given that the dipole amplitude in that energy bin is not significant, with an isotropic probability of $14\%$ \cite{AugerLSA24}.

\begin{figure}[!ht]
    \centering \includegraphics[width=.75\textwidth]{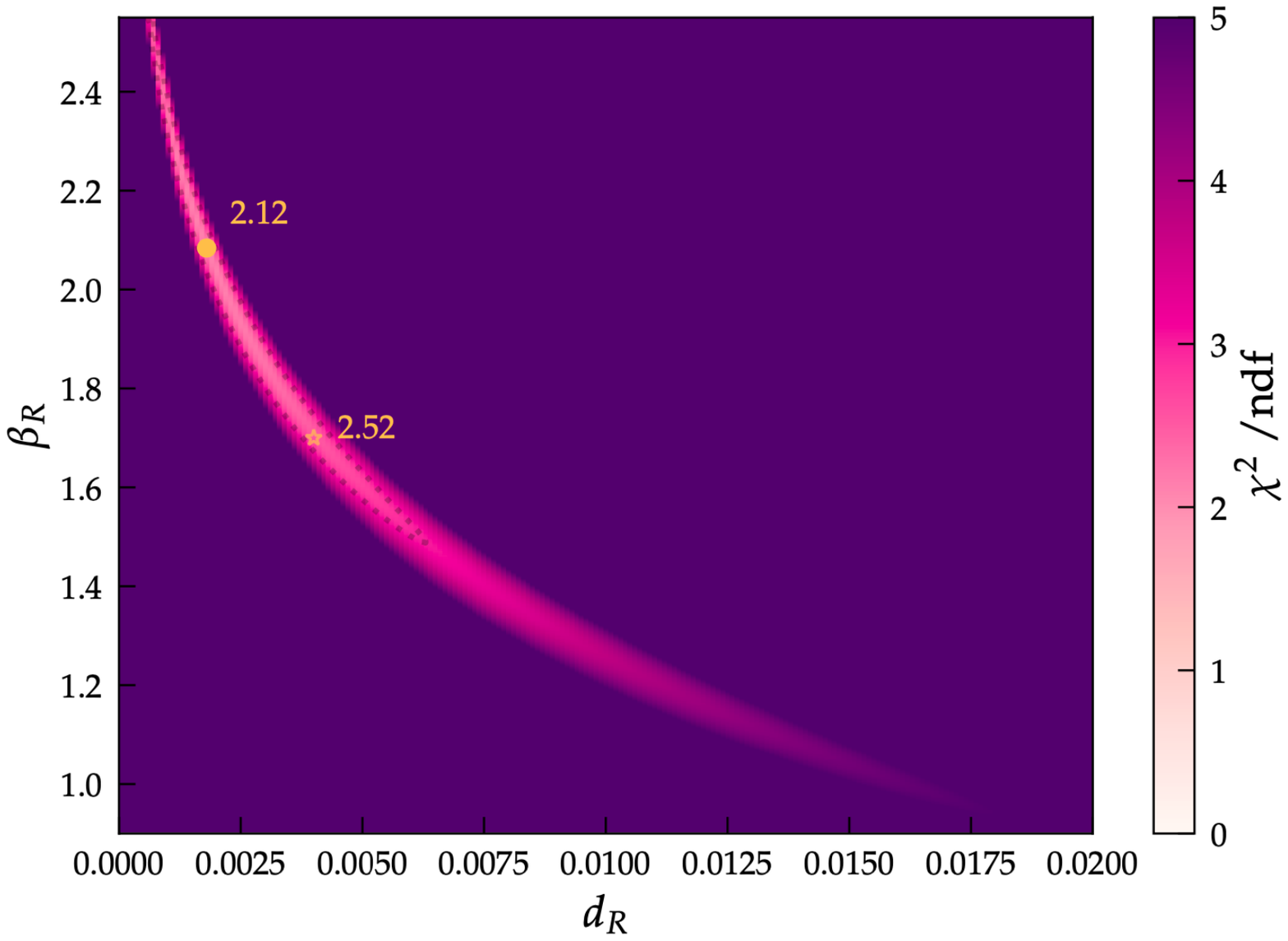}
    \hfill
    \includegraphics[width=.75\textwidth] {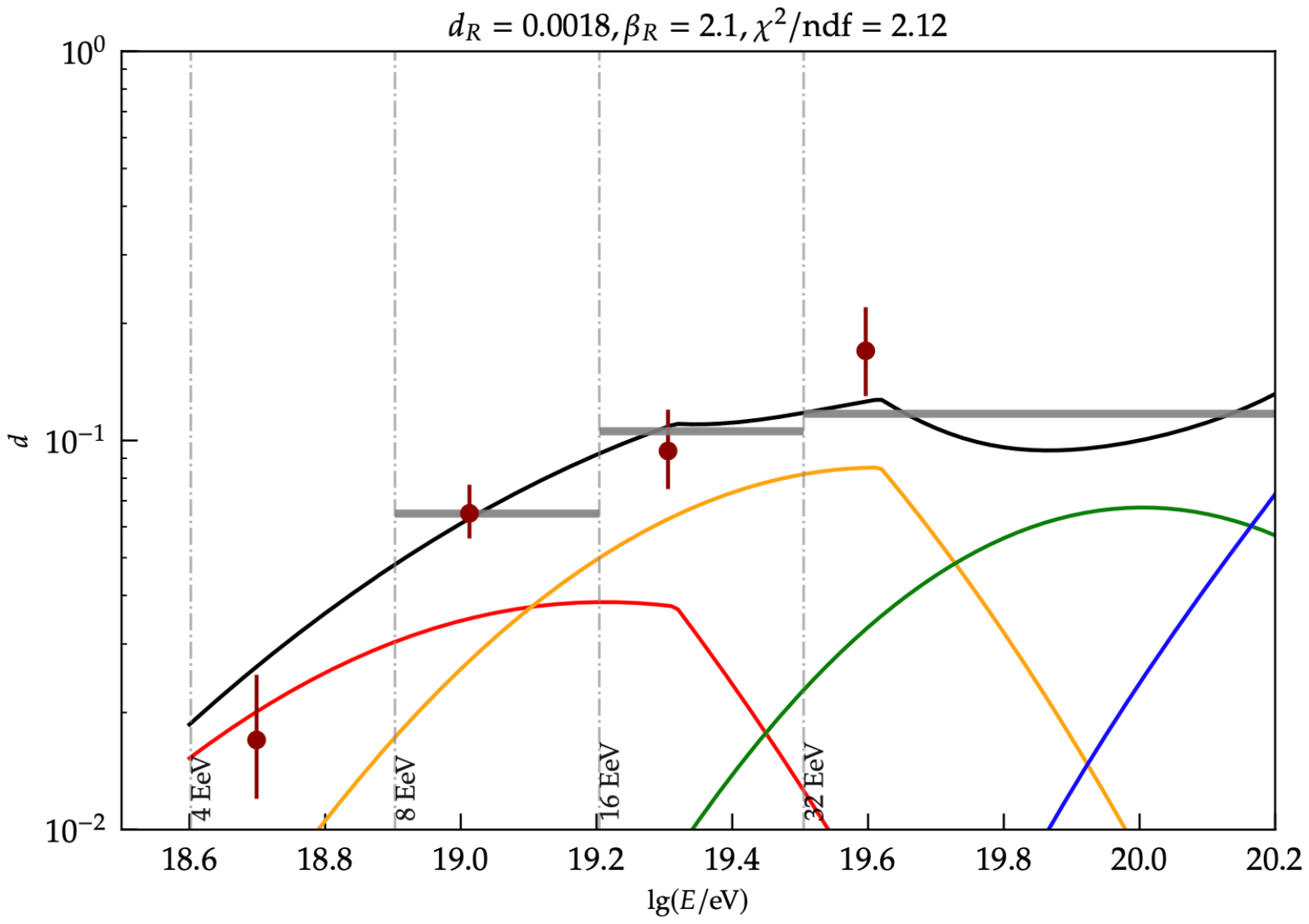}
\caption{Top panel: map of the $\chi^2/{\rm ndf}$ values obtained for the possible values of the parameters $d_R$ and $\beta_R$. The best fit to data is obtained for $d_R=0.0018$ and $\beta_R=2.1$, with a $\chi^2/{\rm ndf=2.12}$. Bottom panel: total dipole amplitude (black) and dipole amplitude predicted for each component following Eq. (1), for proton (red), helium (orange), nitrogen (green), and iron (blue). The gray bars are used to indicate the total dipole amplitude in the energy bins: (8-16, 16-32, $\ge32$)\,EeV. The dipole amplitudes obtained in data are shown with brown circles.}
    \label{fig:Parameters}
\end{figure}

\section{Discovery potential}

The ``lighter'' and ``heavier'' subsets are defined by maximizing the dipole amplitude difference between the two populations, for each energy bin given that the mass composition changes with energy. This difference is computed with the standardized mean difference (SMD),
\begin{equation}
{\rm SMD}=\frac{|d_{\rm light}-d_{\rm heavy}|}{\sqrt{\sigma^2_{\rm light}+\sigma^2_{\rm heavy}}}
\end{equation}
with $\sigma_{\rm light}, \sigma_{\rm heavy}$ the statistical uncertainty of the dipolar amplitudes of each population, $d_{\rm light},d_{\rm heavy}$, computed as $\sigma=\sqrt{2/N}$, with $N$ the total number of events for each population in the corresponding energy bin. 

The ``light'' and ``heavy'' populations are defined for each energy bin based on the tails of the distribution of each mass estimator, with a subset of intermediate events excluded from either category. For the mass estimator based on air-shower universality, the ``light'' population is defined as all the events with $\ln A^{\rm Univ}$ below the $\ln A_{{\rm light}}^{\rm Univ}$ threshold, while the ``heavy'' population consists of the events with $\ln A^{\rm Univ} >\ln A_{{\rm heavy}}^{\rm Univ}$. Similarly, for the mass estimator derived with deep learning, we define the classification threshold based on $X_{{\rm max}}^{19, {\rm DNN}}$, which represents the reconstructed depth of the shower maximum at $10^{19}\,$eV corrected for energy dependence as $X_{{\rm max}}^{19, {\rm DNN}}=X_{{\rm max}}^{{\rm DNN}}- 58\, {\rm g\, cm}^{-2} \log(E/10^{19}\,{\rm eV})$. Accordingly, the ``light'' and ``heavy'' populations are those events with $X_{{\rm max}}^{19, {\rm DNN}}$ above $X_{{\rm max, light}}^{19, {\rm DNN}}$ and below $X_{{\rm max, heavy}}^{19, {\rm DNN}}$, respectively.

Given that the composition fractions for the different elements vary with energy, the thresholds are defined separately for each energy bin. Thus, what is considered to be ``light'' or ``heavy'' is relative to the mass composition distribution within that energy bin, with thresholds optimized to maximize the dipole amplitude difference. For example, in the hypothetical case where there were an energy bin containing only nitrogen and iron, the ``light'' population would correspond to nitrogen, and the ``heavy'' population to iron.

The optimal thresholds that maximize the dipole amplitude difference between the ``light'' and ``heavy'' populations for the two mass estimators considered, for the energy bins of (8-16, 16-32, $\ge32$)\,EeV, are summarized in Table 1. The fraction of events — relative to the total number of events in each energy bin, $N_{\rm tot}$ — that belong to each population are also reported. The optimal thresholds for $\ln A^{\rm Univ}$ increase with energy, as expected from the composition model assumed, based on the results on the mass composition fractions. Analogously, the optimal thresholds $X_{{\rm max, th}}^{19, {\rm DNN}}$ decrease with energy. As a reminder, $\ln A^{\rm Univ}$ serves as a proxy for the physical $\ln A$, which is why $\ln A^{\rm Univ}$ can adopt negative values.  

In Fig.~4 we show the expected dipole amplitudes as a function of energy for the ``light'' and ``heavy'' populations, defined using the thresholds presented in Table 1. A clear separation is observed for both mass estimators. The SMD values obtained for the Universality and DNN-based mass estimators, for each energy bin, are included in Table 1. From the SMD values we conclude that, for the same total number of events, the DNN-based mass estimator, which has a better $X_{{\rm max}}$ resolution than the Universality-based one, is expected to lead to less mixing between the different mass elements and allows for a larger dipole difference between ``light'' and ``heavy'' populations.

One should note that, this model is source-agnostic given that the same dipole direction was assumed for both populations. Allowing for different directions could further enhance the vector separation between them.

Obtaining these cuts for the ``light'' and ``heavy'' populations using simulations that follow the Auger mass composition and spectrum allows us to apply them in data without an additional penalization to the significance of the results. Such penalization would be necessary if the cut parameters were instead optimized directly on the data.

It is important to note that when applying this analysis to data, spurious modulations in the mass estimators are expected due to the influence of atmospheric variations and the geomagnetic field on air-shower development. These modulations should be corrected, similar to the corrections applied to the event energy for weather and geomagnetic effects.

\begin{figure}[!ht]
    \centering \includegraphics[width=.8\textwidth]{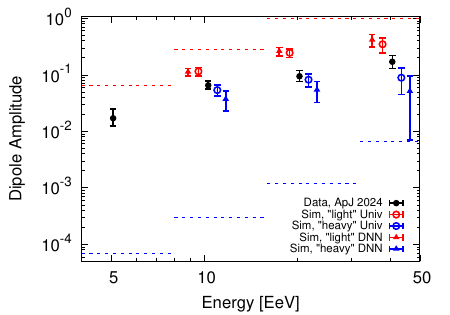}
\caption{Expected dipole amplitude as a function of energy for the ``light'' (red) and ``heavy'' (blue) populations using mass estimators reconstructed with air-shower universality (Univ, empty circles) and deep learning (DNN, filled triangles). The black circles are the dipole amplitudes from data and the red (blue) dashed lines are the predicted amplitudes for proton (iron) with the model assumed in this work.}
    \label{fig:Disco}
\end{figure}

\begin{table}[h]
    \centering
    \begin{tabular}{c|c|c|ccc|ccc}
    \hline \hline
	& & & \multicolumn{3}{c|}{${\rm Universality}$} & \multicolumn{3}{c}{${\rm DNN}$}\\
  	 $E$ & $N_{\rm tot}$ & Population & $\ln A_{{\rm th}}^{\rm Univ}$ & fraction  & SMD & $X_{{\rm max, th}}^{19, {\rm DNN}}$ & fraction & SMD \\
     
  	 [EeV] & & & &  [\%] & &  [g cm$^{-2}]$ &  [\%] &	\\    
	\hline
    \multirow{2}{*}{[8-16]} & \multirow{2}{*}{$28612$} & ``light'' & $-0.6$ & $19.5$ & \multirow{2}{*}{$2.7$} & $793$ & $21.9$ & \multirow{2}{*}{$3.3$} \\  
           & &``heavy''  & $\phantom{+}2.0$  & $50.0$ & & $741$ & $36.8$ &  \\
    \hline
    \multirow{2}{*}{[16-32]}& \multirow{2}{*}{$8024$} & ``light'' & $-0.2$ & $13.4$ & \multirow{2}{*}{$3.4$} & $790$ & $14.6$ & \multirow{2}{*}{$4.3$}\\  
           & &``heavy''& $\phantom{+}2.2$  & $57.0$ & & $740$ & $48.4$ & \\
    \hline
     \multirow{2}{*}{>32}   & \multirow{2}{*}{$2136$} & ``light''  & $\phantom{+}0.0$ & $\phantom{0}8.7$ & \multirow{2}{*}{$2.3$} &$781$ & $\phantom{0}9.1$ & \multirow{2}{*}{$3.2$} \\ 
           & &``heavy''  & $\phantom{+}3.4$  & $46.3$ & & $722$ & $47.7$ & \\
    \hline
    \end{tabular}
    \caption{Optimal thresholds, $\ln A_{{\rm th}}^{\rm Univ}$ and $X_{{\rm max, th}}^{19, {\rm DNN}}$, and fractions of events for the ``light'' and ``heavy'' populations, for each energy bin, using mass estimators reconstructed with Universality and DNN, respectively. The total number of events, $N_{\rm tot}$, considered in this work, for each energy bin and the SMD values obtained are also included.}
\end{table}

\section{Conclusions and outlook}
We investigated the rigidity dependence of the dipole amplitude for the three energy bins above $8\,$EeV by dividing the events into composition-distinct subsets --``light'' and ``heavy''-- based on a composition model that reproduces measurements with the Auger Observatory and extrapolates them to higher energies. Using simulations, we assessed the feasibility of detecting a separation in the total dipole amplitude between these two populations in each energy bin, employing two different mass estimators derived from the surface detector data. The results indicate a positive prospect for such a detection. When applying this model to data, it is necessary to correct for spurious modulations in $X_{\rm max}$ caused by weather and geomagnetic effects, similar to the energy corrections performed in standard large-scale anisotropy analyses. The implementation of this analysis on data is currently in progress.

\clearpage
\section*{The Pierre Auger Collaboration}

{\footnotesize\setlength{\baselineskip}{10pt}
\noindent
\begin{wrapfigure}[11]{l}{0.12\linewidth}
\vspace{-4pt}
\includegraphics[width=0.98\linewidth]{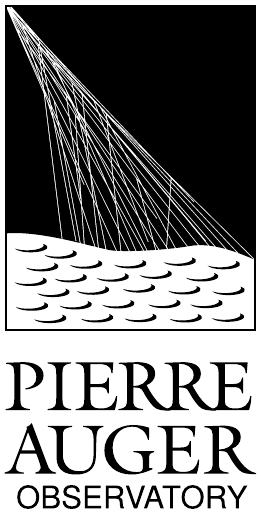}
\end{wrapfigure}
\begin{sloppypar}\noindent
A.~Abdul Halim$^{13}$,
P.~Abreu$^{70}$,
M.~Aglietta$^{53,51}$,
I.~Allekotte$^{1}$,
K.~Almeida Cheminant$^{78,77}$,
A.~Almela$^{7,12}$,
R.~Aloisio$^{44,45}$,
J.~Alvarez-Mu\~niz$^{76}$,
A.~Ambrosone$^{44}$,
J.~Ammerman Yebra$^{76}$
% ...

\end{sloppypar}
\begin{center}
\end{center}

\vspace{1ex}
\begin{description}[labelsep=0.2em,align=right,labelwidth=0.7em,labelindent=0em,leftmargin=2em,noitemsep,before={\renewcommand\makelabel[1]{##1 }}]
\item[$^{1}$] Centro At\'omico Bariloche and Instituto Balseiro (CNEA-UNCuyo-CONICET), San Carlos de Bariloche, Argentina
\item[$^{2}$] Departamento de F\'\i{}sica and Departamento de Ciencias de la Atm\'osfera y los Oc\'eanos, FCEyN, Universidad de Buenos Aires and CONICET, Buenos Aires, Argentina
\item[$^{3}$] IFLP, Universidad Nacional de La Plata and CONICET, La Plata, Argentina
\item[$^{4}$] Instituto de Astronom\'\i{}a y F\'\i{}sica del Espacio (IAFE, CONICET-UBA), Buenos Aires, Argentina
\item[$^{5}$] Instituto de F\'\i{}sica de Rosario (IFIR) -- CONICET/U.N.R.\ and Facultad de Ciencias Bioqu\'\i{}micas y Farmac\'euticas U.N.R., Rosario, Argentina
\item[$^{6}$] Instituto de Tecnolog\'\i{}as en Detecci\'on y Astropart\'\i{}culas (CNEA, CONICET, UNSAM), and Universidad Tecnol\'ogica Nacional -- Facultad Regional Mendoza (CONICET/CNEA), Mendoza, Argentina
\item[$^{7}$] Instituto de Tecnolog\'\i{}as en Detecci\'on y Astropart\'\i{}culas (CNEA, CONICET, UNSAM), Buenos Aires, Argentina
\item[$^{8}$] International Center of Advanced Studies and Instituto de Ciencias F\'\i{}sicas, ECyT-UNSAM and CONICET, Campus Miguelete -- San Mart\'\i{}n, Buenos Aires, Argentina
\item[$^{9}$] Laboratorio Atm\'osfera -- Departamento de Investigaciones en L\'aseres y sus Aplicaciones -- UNIDEF (CITEDEF-CONICET), Argentina
\item[$^{10}$] Observatorio Pierre Auger, Malarg\"ue, Argentina
\item[$^{11}$] Observatorio Pierre Auger and Comisi\'on Nacional de Energ\'\i{}a At\'omica, Malarg\"ue, Argentina
\item[$^{12}$] Universidad Tecnol\'ogica Nacional -- Facultad Regional Buenos Aires, Buenos Aires, Argentina
\item[$^{13}$] University of Adelaide, Adelaide, S.A., Australia
\item[$^{14}$] Universit\'e Libre de Bruxelles (ULB), Brussels, Belgium
\item[$^{15}$] Vrije Universiteit Brussels, Brussels, Belgium
\item[$^{16}$] Centro Brasileiro de Pesquisas Fisicas, Rio de Janeiro, RJ, Brazil
\item[$^{17}$] Centro Federal de Educa\c{c}\~ao Tecnol\'ogica Celso Suckow da Fonseca, Petropolis, Brazil
\item[$^{18}$] Instituto Federal de Educa\c{c}\~ao, Ci\^encia e Tecnologia do Rio de Janeiro (IFRJ), Brazil
\item[$^{19}$] Universidade de S\~ao Paulo, Escola de Engenharia de Lorena, Lorena, SP, Brazil
\item[$^{20}$] Universidade de S\~ao Paulo, Instituto de F\'\i{}sica de S\~ao Carlos, S\~ao Carlos, SP, Brazil
\item[$^{21}$] Universidade de S\~ao Paulo, Instituto de F\'\i{}sica, S\~ao Paulo, SP, Brazil
\item[$^{22}$] Universidade Estadual de Campinas (UNICAMP), IFGW, Campinas, SP, Brazil
\item[$^{23}$] Universidade Estadual de Feira de Santana, Feira de Santana, Brazil
\item[$^{24}$] Universidade Federal de Campina Grande, Centro de Ciencias e Tecnologia, Campina Grande, Brazil
\item[$^{25}$] Universidade Federal do ABC, Santo Andr\'e, SP, Brazil
\item[$^{26}$] Universidade Federal do Paran\'a, Setor Palotina, Palotina, Brazil
\item[$^{27}$] Universidade Federal do Rio de Janeiro, Instituto de F\'\i{}sica, Rio de Janeiro, RJ, Brazil
\item[$^{28}$] Universidad de Medell\'\i{}n, Medell\'\i{}n, Colombia
\item[$^{29}$] Universidad Industrial de Santander, Bucaramanga, Colombia
\item[$^{30}$] Charles University, Faculty of Mathematics and Physics, Institute of Particle and Nuclear Physics, Prague, Czech Republic
\item[$^{31}$] Institute of Physics of the Czech Academy of Sciences, Prague, Czech Republic
\item[$^{32}$] Palacky University, Olomouc, Czech Republic
\item[$^{33}$] CNRS/IN2P3, IJCLab, Universit\'e Paris-Saclay, Orsay, France
\item[$^{34}$] Laboratoire de Physique Nucl\'eaire et de Hautes Energies (LPNHE), Sorbonne Universit\'e, Universit\'e de Paris, CNRS-IN2P3, Paris, France
\item[$^{35}$] Univ.\ Grenoble Alpes, CNRS, Grenoble Institute of Engineering Univ.\ Grenoble Alpes, LPSC-IN2P3, 38000 Grenoble, France
\item[$^{36}$] Universit\'e Paris-Saclay, CNRS/IN2P3, IJCLab, Orsay, France
\item[$^{37}$] Bergische Universit\"at Wuppertal, Department of Physics, Wuppertal, Germany
\item[$^{38}$] Karlsruhe Institute of Technology (KIT), Institute for Experimental Particle Physics, Karlsruhe, Germany
\item[$^{39}$] Karlsruhe Institute of Technology (KIT), Institut f\"ur Prozessdatenverarbeitung und Elektronik, Karlsruhe, Germany
\item[$^{40}$] Karlsruhe Institute of Technology (KIT), Institute for Astroparticle Physics, Karlsruhe, Germany
\item[$^{41}$] RWTH Aachen University, III.\ Physikalisches Institut A, Aachen, Germany
\item[$^{42}$] Universit\"at Hamburg, II.\ Institut f\"ur Theoretische Physik, Hamburg, Germany
\item[$^{43}$] Universit\"at Siegen, Department Physik -- Experimentelle Teilchenphysik, Siegen, Germany
\item[$^{44}$] Gran Sasso Science Institute, L'Aquila, Italy
\item[$^{45}$] INFN Laboratori Nazionali del Gran Sasso, Assergi (L'Aquila), Italy
\item[$^{46}$] INFN, Sezione di Catania, Catania, Italy
\item[$^{47}$] INFN, Sezione di Lecce, Lecce, Italy
\item[$^{48}$] INFN, Sezione di Milano, Milano, Italy
\item[$^{49}$] INFN, Sezione di Napoli, Napoli, Italy
\item[$^{50}$] INFN, Sezione di Roma ``Tor Vergata'', Roma, Italy
\item[$^{51}$] INFN, Sezione di Torino, Torino, Italy
\item[$^{52}$] Istituto di Astrofisica Spaziale e Fisica Cosmica di Palermo (INAF), Palermo, Italy
\item[$^{53}$] Osservatorio Astrofisico di Torino (INAF), Torino, Italy
\item[$^{54}$] Politecnico di Milano, Dipartimento di Scienze e Tecnologie Aerospaziali , Milano, Italy
\item[$^{55}$] Universit\`a del Salento, Dipartimento di Matematica e Fisica ``E.\ De Giorgi'', Lecce, Italy
\item[$^{56}$] Universit\`a dell'Aquila, Dipartimento di Scienze Fisiche e Chimiche, L'Aquila, Italy
\item[$^{57}$] Universit\`a di Catania, Dipartimento di Fisica e Astronomia ``Ettore Majorana``, Catania, Italy
\item[$^{58}$] Universit\`a di Milano, Dipartimento di Fisica, Milano, Italy
\item[$^{59}$] Universit\`a di Napoli ``Federico II'', Dipartimento di Fisica ``Ettore Pancini'', Napoli, Italy
\item[$^{60}$] Universit\`a di Palermo, Dipartimento di Fisica e Chimica ''E.\ Segr\`e'', Palermo, Italy
\item[$^{61}$] Universit\`a di Roma ``Tor Vergata'', Dipartimento di Fisica, Roma, Italy
\item[$^{62}$] Universit\`a Torino, Dipartimento di Fisica, Torino, Italy
\item[$^{63}$] Benem\'erita Universidad Aut\'onoma de Puebla, Puebla, M\'exico
\item[$^{64}$] Unidad Profesional Interdisciplinaria en Ingenier\'\i{}a y Tecnolog\'\i{}as Avanzadas del Instituto Polit\'ecnico Nacional (UPIITA-IPN), M\'exico, D.F., M\'exico
\item[$^{65}$] Universidad Aut\'onoma de Chiapas, Tuxtla Guti\'errez, Chiapas, M\'exico
\item[$^{66}$] Universidad Michoacana de San Nicol\'as de Hidalgo, Morelia, Michoac\'an, M\'exico
\item[$^{67}$] Universidad Nacional Aut\'onoma de M\'exico, M\'exico, D.F., M\'exico
\item[$^{68}$] Institute of Nuclear Physics PAN, Krakow, Poland
\item[$^{69}$] University of \L{}\'od\'z, Faculty of High-Energy Astrophysics,\L{}\'od\'z, Poland
\item[$^{70}$] Laborat\'orio de Instrumenta\c{c}\~ao e F\'\i{}sica Experimental de Part\'\i{}culas -- LIP and Instituto Superior T\'ecnico -- IST, Universidade de Lisboa -- UL, Lisboa, Portugal
\item[$^{71}$] ``Horia Hulubei'' National Institute for Physics and Nuclear Engineering, Bucharest-Magurele, Romania
\item[$^{72}$] Institute of Space Science, Bucharest-Magurele, Romania
\item[$^{73}$] Center for Astrophysics and Cosmology (CAC), University of Nova Gorica, Nova Gorica, Slovenia
\item[$^{74}$] Experimental Particle Physics Department, J.\ Stefan Institute, Ljubljana, Slovenia
\item[$^{75}$] Universidad de Granada and C.A.F.P.E., Granada, Spain
\item[$^{76}$] Instituto Galego de F\'\i{}sica de Altas Enerx\'\i{}as (IGFAE), Universidade de Santiago de Compostela, Santiago de Compostela, Spain
\item[$^{77}$] IMAPP, Radboud University Nijmegen, Nijmegen, The Netherlands
\item[$^{78}$] Nationaal Instituut voor Kernfysica en Hoge Energie Fysica (NIKHEF), Science Park, Amsterdam, The Netherlands
\item[$^{79}$] Stichting Astronomisch Onderzoek in Nederland (ASTRON), Dwingeloo, The Netherlands
\item[$^{80}$] Universiteit van Amsterdam, Faculty of Science, Amsterdam, The Netherlands
\item[$^{81}$] Case Western Reserve University, Cleveland, OH, USA
\item[$^{82}$] Colorado School of Mines, Golden, CO, USA
\item[$^{83}$] Department of Physics and Astronomy, Lehman College, City University of New York, Bronx, NY, USA
\item[$^{84}$] Michigan Technological University, Houghton, MI, USA
\item[$^{85}$] New York University, New York, NY, USA
\item[$^{86}$] University of Chicago, Enrico Fermi Institute, Chicago, IL, USA
\item[$^{87}$] University of Delaware, Department of Physics and Astronomy, Bartol Research Institute, Newark, DE, USA
\item[] -----
\item[$^{a}$] Max-Planck-Institut f\"ur Radioastronomie, Bonn, Germany
\item[$^{b}$] also at Kapteyn Institute, University of Groningen, Groningen, The Netherlands
\item[$^{c}$] School of Physics and Astronomy, University of Leeds, Leeds, United Kingdom
\item[$^{d}$] Fermi National Accelerator Laboratory, Fermilab, Batavia, IL, USA
\item[$^{e}$] Pennsylvania State University, University Park, PA, USA
\item[$^{f}$] Colorado State University, Fort Collins, CO, USA
\item[$^{g}$] Louisiana State University, Baton Rouge, LA, USA
\item[$^{h}$] now at Graduate School of Science, Osaka Metropolitan University, Osaka, Japan
\item[$^{i}$] Institut universitaire de France (IUF), France
\item[$^{j}$] now at Technische Universit\"at Dortmund and Ruhr-Universit\"at Bochum, Dortmund and Bochum, Germany
\end{description}

\section*{Acknowledgments}

\begin{sloppypar}
The successful installation, commissioning, and operation of the Pierre
Auger Observatory would not have been possible without the strong
commitment and effort from the technical and administrative staff in
Malarg\"ue. We are very grateful to the following agencies and
organizations for financial support:
\end{sloppypar}

\begin{sloppypar}
Argentina -- Comisi\'on Nacional de Energ\'\i{}a At\'omica; Agencia Nacional de
Promoci\'on Cient\'\i{}fica y Tecnol\'ogica (ANPCyT); Consejo Nacional de
Investigaciones Cient\'\i{}ficas y T\'ecnicas (CONICET); Gobierno de la
Provincia de Mendoza; Municipalidad de Malarg\"ue; NDM Holdings and Valle
Las Le\~nas; in gratitude for their continuing cooperation over land
access; Australia -- the Australian Research Council; Belgium -- Fonds
de la Recherche Scientifique (FNRS); Research Foundation Flanders (FWO),
Marie Curie Action of the European Union Grant No.~101107047; Brazil --
Conselho Nacional de Desenvolvimento Cient\'\i{}fico e Tecnol\'ogico (CNPq);
Financiadora de Estudos e Projetos (FINEP); Funda\c{c}\~ao de Amparo \`a
Pesquisa do Estado de Rio de Janeiro (FAPERJ); S\~ao Paulo Research
Foundation (FAPESP) Grants No.~2019/10151-2, No.~2010/07359-6 and
No.~1999/05404-3; Minist\'erio da Ci\^encia, Tecnologia, Inova\c{c}\~oes e
Comunica\c{c}\~oes (MCTIC); Czech Republic -- GACR 24-13049S, CAS LQ100102401,
MEYS LM2023032, CZ.02.1.01/0.0/0.0/16{\textunderscore}013/0001402,
CZ.02.1.01/0.0/0.0/18{\textunderscore}046/0016010 and
CZ.02.1.01/0.0/0.0/17{\textunderscore}049/0008422 and CZ.02.01.01/00/22{\textunderscore}008/0004632;
France -- Centre de Calcul IN2P3/CNRS; Centre National de la Recherche
Scientifique (CNRS); Conseil R\'egional Ile-de-France; D\'epartement
Physique Nucl\'eaire et Corpusculaire (PNC-IN2P3/CNRS); D\'epartement
Sciences de l'Univers (SDU-INSU/CNRS); Institut Lagrange de Paris (ILP)
Grant No.~LABEX ANR-10-LABX-63 within the Investissements d'Avenir
Programme Grant No.~ANR-11-IDEX-0004-02; Germany -- Bundesministerium
f\"ur Bildung und Forschung (BMBF); Deutsche Forschungsgemeinschaft (DFG);
Finanzministerium Baden-W\"urttemberg; Helmholtz Alliance for
Astroparticle Physics (HAP); Helmholtz-Gemeinschaft Deutscher
Forschungszentren (HGF); Ministerium f\"ur Kultur und Wissenschaft des
Landes Nordrhein-Westfalen; Ministerium f\"ur Wissenschaft, Forschung und
Kunst des Landes Baden-W\"urttemberg; Italy -- Istituto Nazionale di
Fisica Nucleare (INFN); Istituto Nazionale di Astrofisica (INAF);
Ministero dell'Universit\`a e della Ricerca (MUR); CETEMPS Center of
Excellence; Ministero degli Affari Esteri (MAE), ICSC Centro Nazionale
di Ricerca in High Performance Computing, Big Data and Quantum
Computing, funded by European Union NextGenerationEU, reference code
CN{\textunderscore}00000013; M\'exico -- Consejo Nacional de Ciencia y Tecnolog\'\i{}a
(CONACYT) No.~167733; Universidad Nacional Aut\'onoma de M\'exico (UNAM);
PAPIIT DGAPA-UNAM; The Netherlands -- Ministry of Education, Culture and
Science; Netherlands Organisation for Scientific Research (NWO); Dutch
national e-infrastructure with the support of SURF Cooperative; Poland
-- Ministry of Education and Science, grants No.~DIR/WK/2018/11 and
2022/WK/12; National Science Centre, grants No.~2016/22/M/ST9/00198,
2016/23/B/ST9/01635, 2020/39/B/ST9/01398, and 2022/45/B/ST9/02163;
Portugal -- Portuguese national funds and FEDER funds within Programa
Operacional Factores de Competitividade through Funda\c{c}\~ao para a Ci\^encia
e a Tecnologia (COMPETE); Romania -- Ministry of Research, Innovation
and Digitization, CNCS-UEFISCDI, contract no.~30N/2023 under Romanian
National Core Program LAPLAS VII, grant no.~PN 23 21 01 02 and project
number PN-III-P1-1.1-TE-2021-0924/TE57/2022, within PNCDI III; Slovenia
-- Slovenian Research Agency, grants P1-0031, P1-0385, I0-0033, N1-0111;
Spain -- Ministerio de Ciencia e Innovaci\'on/Agencia Estatal de
Investigaci\'on (PID2019-105544GB-I00, PID2022-140510NB-I00 and
RYC2019-027017-I), Xunta de Galicia (CIGUS Network of Research Centers,
Consolidaci\'on 2021 GRC GI-2033, ED431C-2021/22 and ED431F-2022/15),
Junta de Andaluc\'\i{}a (SOMM17/6104/UGR and P18-FR-4314), and the European
Union (Marie Sklodowska-Curie 101065027 and ERDF); USA -- Department of
Energy, Contracts No.~DE-AC02-07CH11359, No.~DE-FR02-04ER41300,
No.~DE-FG02-99ER41107 and No.~DE-SC0011689; National Science Foundation,
Grant No.~0450696, and NSF-2013199; The Grainger Foundation; Marie
Curie-IRSES/EPLANET; European Particle Physics Latin American Network;
and UNESCO.
\end{sloppypar}

}

%
%
%

\end{document}